\documentclass[letterpaper,twocolumn,10pt]{article}
\usepackage{usenix,epsfig}
\usepackage{url}
\usepackage{todonotes}
\usepackage{verbatim}
\usepackage{listings}
\usepackage[T1]{fontenc}
\usepackage[utf8]{inputenc}\begin{document}

\date{}

\title{\Large \bf Finding Tizen security bugs through whole-system
  static analysis}

\author{Daniel Song, Jisheng Zhao, Michael Burke, Dragoş Sbîrlea, Dan Wallach, and Vivek Sarkar\\Rice University\\ \{dwsong, jisheng.zhao, mgb2, dragos, dwallach, vsarkar\}@rice.edu}

\maketitle

\thispagestyle{empty}

\subsection*{Abstract}
Tizen is a new Linux-based open source platform for consumer devices
including smartphones, televisions, vehicles, and wearables. While
Tizen provides kernel-level mandatory policy enforcement, it has a
large collection of libraries, implemented in a mix of C and C++,
which make their own security checks. In this research, we describe
the design and engineering of a static analysis engine which drives a full 
information flow analysis for apps and a control flow analysis for the
full library stack. We implemented these static analyses as extensions
to LLVM, requiring us to improve LLVM's native analysis features to
get greater precision and scalability, including knotty issues like
the coexistence of C++ inheritance with C function pointer use. With
our tools, we found several unexpected behaviors in the Tizen system,
including paths through the system libraries that did not have inline
security checks. We show how our tools can help the Tizen app
store to verify important app properties as well as helping the Tizen
development process avoid the accidental introduction of subtle
vulnerabilities.

\section{Introduction}\label{sec:intro}
Static analysis has proven to be wildly successful in finding all sorts of bugs, whether related to security or other flaws, so the availability of a new system to analyze for bugs is an interesting opportunity to see how good these tools can be. To that end, we had the opportunity to design and implement static analyses for Tizen, a new operating system platform that will soon run on a variety of Samsung products including televisions, wearables, automobile telematics systems, and smartphones. This paper describes the analysis challenges presented by the Tizen platform, as distinct from competing  platforms like Android, along with the tools we developed and the issues we found. 

We'll describe the Tizen architecture in more detail later, but at a high level Tizen is a variant of Linux, with kernel-enforced mandatory access control rules. Applications can be built entirely from HTML5 web primitives (JavaScript, etc.), much as was done in Palm's WebOS, or they can be built natively, using a variety of C and C++ standard libraries. Tizen has a series of permissions that can be granted to applications in a fashion similar to Android, which are then enforced both at a low-level, using the kernel, along with higher-level checks embedded in the libraries. Native apps will be distributed as LLVM bitcode---a portable, machine-independent intermediate code representation that's naturally amenable to static analysis via the LLVM toolchain. We presume there will be a centralized Tizen app store---Samsung just opened TizenStore.com in January of this year---that can conduct analyses over Tizen apps to ensure their safety prior to being downloaded to Tizen users\footnote{While the authors of this paper are blinded for review, we note that we do not represent Samsung, Intel, or any other commercial company involved in Tizen.  All the work presented here is based on public information including Samsung's open-source release of the Tizen codebase.}. In a recent talk, Samsung's partner, AhnLabs, described a mixed process with both static and dynamic analysis as well as human analysts~\cite{Ahnlab,TizenValidationDoc}.

In deciding what aspects of the Tizen system were interesting for a security-related static analyses, we decided to focus our attention on higher-level security topics. For native Tizen apps, we concluded that it would be most helpful to have a general-purpose LLVM information flow analysis tool that could identify apps containing undesired flows, such as from the user's contacts list to the network. We envision this automated analysis being conducted mechanically in an app store alongside a human analyst who studies the effectiveness of various source/sink pairs, amending the rules as needed. The goals of this tool are to run quickly and to produce useful evidence that can quickly allow safe apps to be approved, allowing human analysts to spend more of their time digging into suspicious apps with unusual behaviors. We prefer information flow analyses over more primitive cataloging of privileged operations, as done in the Tizen store presently~\cite{Ahnlab,TizenValidationDoc}, because we hypothesize it will result in fewer false positives. For example, if a privacy-sensitive advertising library downloaded several ad variants, selecting one for display based on how well it matches platform-local private information about the user, this would be far less concerning than leaking that same private information over the network for the decision to be made remotely. While both variants use the same permissions, information flow can distinguish the good from the bad.

For the Tizen system libraries, written in a mix of C and C++ and containing internal security checks that make them part of the system's sizable trusted computing base, we face a larger challenge. These libraries enforce security properties while they are simultaneously linked to the same address space as the potentially hostile apps that call them. We consequently expect that the Tizen app store will need to statically analyze apps to ensure they only branch to approved entry points in the system libraries and that they don't exploit unsafe properties of the C language (e.g., indexing beyond the end of an array, overwriting a function pointer, and branching to a forbidden target). Such ``safety'' analyses are well within the province of existing commercial tools, so we didn't implement them. Furthermore, apps built using the web stack (JavaScript, etc.) call into the very same libraries, pointing to the importance of validating these entry points' use of security checks.

Consequently, we decided to implement a control flow analysis over the native libraries in order to discover whether there are paths through the libraries that are missing security checks, and thus might indicate exploitable flaws that such a ``safety'' analysis in the app store might otherwise approve. Unlike our information flow analysis for Tizen apps, we envision this Tizen library analysis to be something that can run for hours, if not days, in the service of Tizen system developers' internal bug finding. Likewise, we envision that Tizen system developers would be able to add trusted code annotations to inform this analysis, although it's essential that such annotations be few and far between, in order to minimize friction to the adoption of our tool.

The rest of this paper describes Tizen in more detail (Section~\ref{sec:background}), then presents our LLVM-based static analysis engine (Section~\ref{sec:analyses}). We follow with our analysis of Tizen apps (Section~\ref{sec:analysis_app}) and API libraries (Section~\ref{sec:analysis_api}). We discuss pragmatic issues (Section~\ref{sec:pragmatic}). We wrap up with prior work (Section~\ref{sec:related}) and conclusions (Section~\ref{sec:conc}).

\section{Background}\label{sec:background}
The Tizen platform~\cite{tizen} is an operating system based on the Linux kernel and the GNU standard C library. It includes a graphics layer based on the Enlightenment Foundation Libraries and the X Window System.

Tizen already runs on smartphones~\cite{z1}, wearables such as watches~\cite{gear}, cameras~\cite{camera}, vehicle infotainment systems~\cite{auto}, TVs~\cite{tv} and in the future refrigerators, air conditioners and washing machines~\cite{future}. Consequently, its security properties become quite important. 
The Tizen libraries are implemented as a C++ layer of programmer-accessible APIs built on top of a C layer of APIs that are deliberately hidden from the application programmers. The intention is that that application programmers won't deal, for example, with the X Window System, but rather will use Tizen's official graphics APIs.

\subsection{Tizen Applications}

\begin{figure*}
	\centering
	\includegraphics[width=1\textwidth,trim=0cm 0cm 0cm 0cm,clip=true]{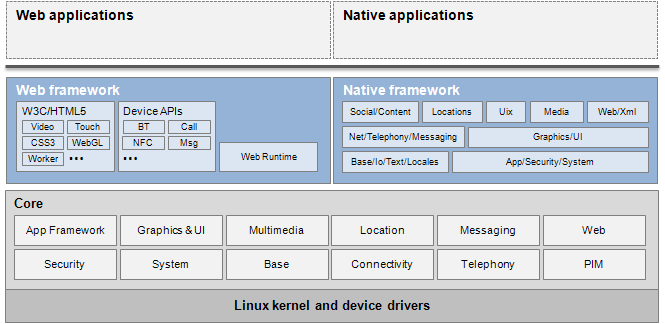}
	\caption{Tizen application development stack. In this paper, we focus on the native applications stack.}
	\label{fig:tizen_layers}
\end{figure*}

Applications can either be based on HTML5 or native apps. This paper focuses on security analysis of the native applications, which use the C standard library and additional Tizen APIs that offer access to phone calling and contacts, SMS, networking, Bluetooth, and other services as shown in Figure ~\ref{fig:tizen_layers}.

The availability and wide range of these APIs  makes Tizen a unique target for analysis since the entirety of kernel, standard libraries, standardized application platform and the applications themselves are written in C/C++ and compiled to native code\footnote{The distribution format for native apps will actually be LLVM's bitcode IR. This intermediate representation, much like Android's Dalvik, will be compiled at install-time to the platform's native CPU architecture.} and are, in effect, all within the trusted computing base of the platform.

\subsection{Tizen Privileges}

The main mechanism for enforcing privacy and security for  applications is a system of privileges, functionally similar to that of Android. The application privileges are displayed to the user ahead of installation, with applications only being downloaded and installed once the user accepts the privileges that the application requires.

From a security standpoint, the use of C/C++ for the Tizen libraries---widely known as difficult to analyze with its use of function pointers, aliased arrays and deep class hierarchies---together with the existence of rich application APIs, each with their own associated permissions, makes determining the correctness of Tizen's privilege system a serious challenge. Even for Android, where privileges are enforced outside of a potentially hostile application's address space, 
researchers have discovered multiple permissions inconsistencies inside the OS libraries~\cite{Felt:11} and several different types of permission misconfiguration~\cite{Sbirlea:13,Felt:11}, leading to application over-privilege~\cite{Wei:12} and increased application vulnerability~\cite{Artz:14,Sellwood:13}.

While, to the best of our knowledge, Tizen does not have a security document explaining the rules of privilege enforcement, by analyzing the code, we observed the following rules.
\begin{itemize}
\item As a first layer of defense, applications are checked for security vulnerabilities before their inclusion in the web store. 

\item Second, an access controller invoked by each privileged  API denies access to the native APIs for which an application does not have the privilege. This is done by including a call to \verb|CheckPrivilege(privilege_name)|. 

\item Third, since checks done in the application process may be avoided by an attacker, protected actions are performed or information is retrieved from other service processes, which perform their own checking for permissions.

\item At the bottom level, the inter-process communication and data access  is protected by a kernel-level security module (SMACK), described below.
\end{itemize}

On its surface, this appears to be an example of {\em defense in depth}, i.e., perhaps the higher-layer checks are unnecessary and SMACK can carry all the security burden, but we hypothesize that the checks at each layer are necessary, as higher-level API semantics may be lost when control flow reaches the system-call boundary. SMACK may not have adequate context to make every security decision correctly on its own.

\subsection{SMACK}

Simplified Mandatory Access Control Kernel (SMACK) is a Linux kernel module and associated utilities that allow setting custom mandatory access control (MAC) rules to protect data and limit process interaction. 

The combination of mandatory access control policies and API privileges for more fine-grained permissions is the standard combination of protection mechanisms in Android, which has its own permissions API and system-level enforcement. Recent versions of Android also include SELinux, which can enforce policies similar to SMACK. 

SMACK relies on labeling system objects and then applying rules, based on those labels, to allow or prevent access. Its rules format is \verb+subject-label object-label access+, where \verb+subject-label+ is the SMACK label of the task, \verb+object-label+ is the SMACK label of the object being accessed, and \verb+access+ is a string specifying the type of access allowed.
We note that the SELinux policy for Linux 2.4.19 consists of over 50,000 policy statements, including over 700 subject types and 100,000 permission assignments~\cite{SELinuxDocs}. While Tizen's SMACK is simpler than SELinux, Tizen 2.1 has 41,000 lines of SMACK access rules~\cite{SmackRules}. It's manifestly unclear whether these rules are ``correct'' or how to even define correctness over them.

\section{Static Analysis Engine}\label{sec:analyses}
\begin{figure}[htbp]
\center
\includegraphics[width=2.8in]{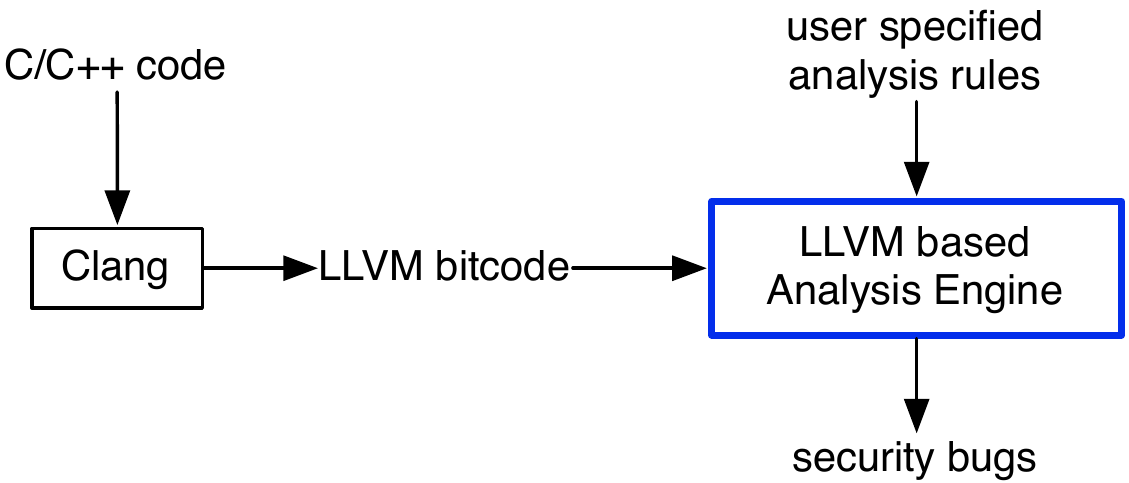}
\caption{The Basic Workflow}
\label{fig:rtas_basic}
\end{figure}

The motivation for this work is to identify security bugs in a C/C++ code base through static analysis.  
The code base could be a mobile application (i.e., a Tizen app) or an operating system (i.e., Tizen). 
We built our analysis infrastructure on top of the LLVM framework.  
Figure ~\ref{fig:rtas_basic} shows the basic flow of our analysis system. The C/C++ code is compiled and
translated to LLVM bitcode by the Clang~\cite{llvm} compiler. The bitcode is input to the LLVM-based analysis engine, 
which performs various information flow-based analyses to identify security bugs. 
The analysis is driven by user-specified {\em analysis rules}, e.g. pairs of taint source and taint sink functions.

This section will describe the analysis engine, including the 
basic components, workflow and the mechanism used to identify two types of security bugs.
The LLVM-based analysis engine applies static analysis to the input bitcode and identifies security 
bugs, i.e. flows that violate the analysis rules.  Section~\ref{sec:analysis_arch} gives an overview of the software architecture 
of the analysis engine. Section~\ref{sec:analysis_basic} describes the static
analysis techniques used in this engine and the interactions among them. 
The last two sections describe how the analysis finds privilege errors and taint pairs in two different kinds of 
code bases: i.e. Tizen applications and the Tizen operating system. 

\subsection{Structure of Analysis Engine}
\label{sec:analysis_arch}
\begin{figure}[htbp]
\center
\includegraphics[width=2.8in]{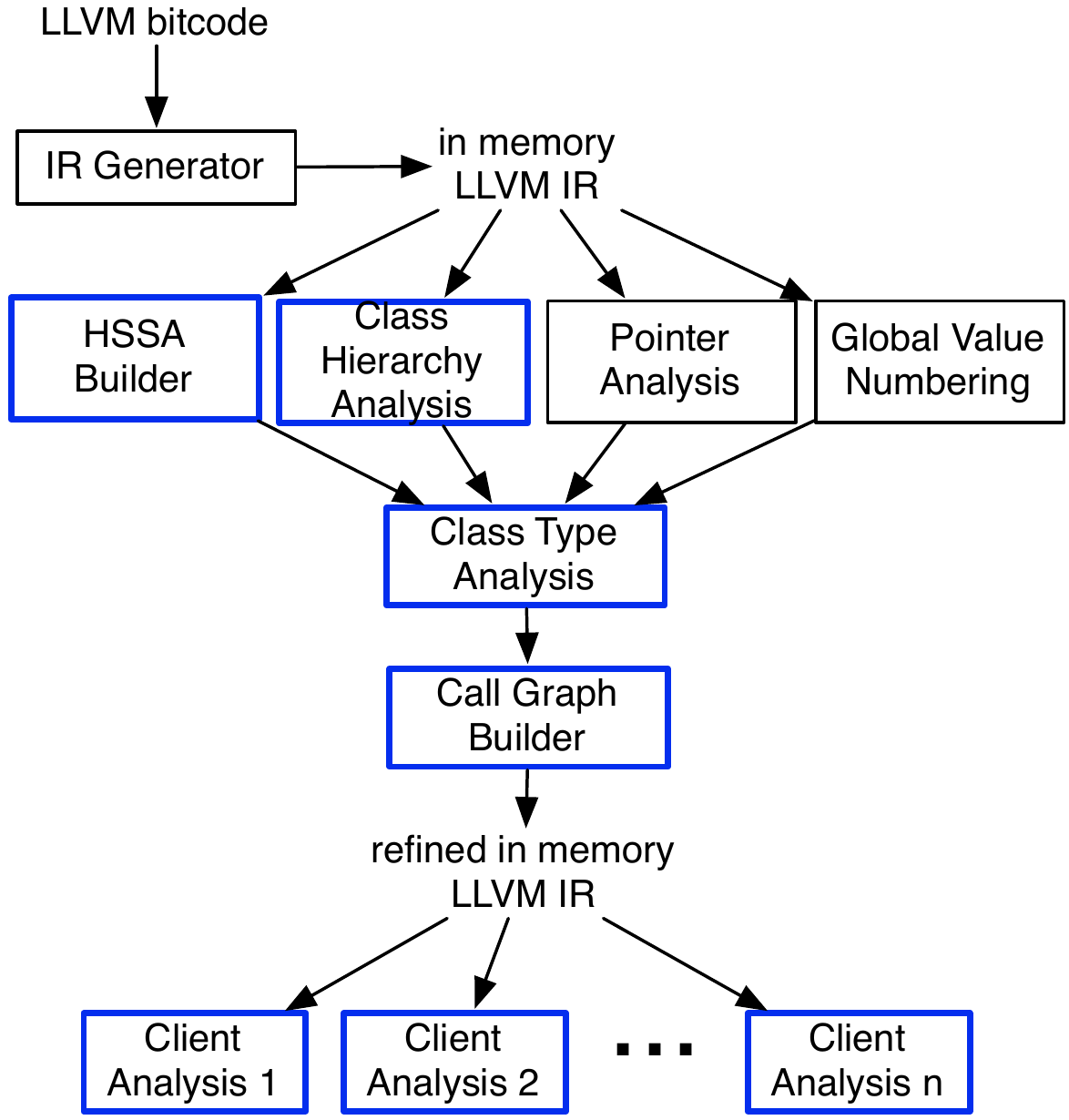}
\caption{The Internal Workflow for LLVM based Analysis Engine}
\label{fig:rtas_analysis}
\end{figure}

Figure~\ref{fig:rtas_analysis} shows the structure of our analysis engine, which is built on the LLVM framework
(the bold boxes indicate components that we have added). The engine takes LLVM bitcode as input and translates
it into an in-memory LLVM intermediate representation (a three-address static-single assignment based IR).
A client analysis is a static information flow analysis (SIFA) that runs on the LLVM IR and identifies
security bugs. To assist the client analysis, a series of auxiliary analyses are invoked to create
additional in-memory information, including the heap static-single assignment (HSSA) form (more detail
is provided in Section~\ref{sec:analysis_basic}), class hierarchy information, class type information
and the call graph. The ``refined in-memory LLVM IR'' is the in-memory LLVM IR augmented by this additional information.

Here we summarize the functionality of the auxiliary analyses \& transformations, and the interactions between them:
\begin{itemize}
\item Class Hierarchy Analysis (CHA): builds the class hierarchy graph for C++ code;
\item HSSA builder: constructs the HSSA form;
\item Pointer Analysis (PTA): intra-procedural pointer analysis;
\item Global Value Numbering (GVN): global value numbering based on PTA;
\item Class Type Analysis (CTA): this is a flow-sensitive class type analysis that is based on CHA, HSSA, PTA and GVN; 
\item Call Graph Builder (CG): the call graph construction based on CTA which can precisely identify
the invoked virtual function calls, including function pointer invocations. 
\end{itemize}

\subsection{Basic Techniques}
\label{sec:analysis_basic} 
This section gives a more detailed description of the functionality of the auxiliary analyses and transformations. 
The pointer analysis (PTA) and global value numbering (GVN) are standard LLVM analysis modules.
The pointer analysis is an intra-procedural stateless analysis that uses allocation sites to
distinguish memory addresses. The global value numbering uses alias information produced by
pointer analysis to number the heap variables that have distinct values. 

\subsubsection{Class Type Analysis}
\begin{figure*}[htbp]
\center
\includegraphics[width=6.8in]{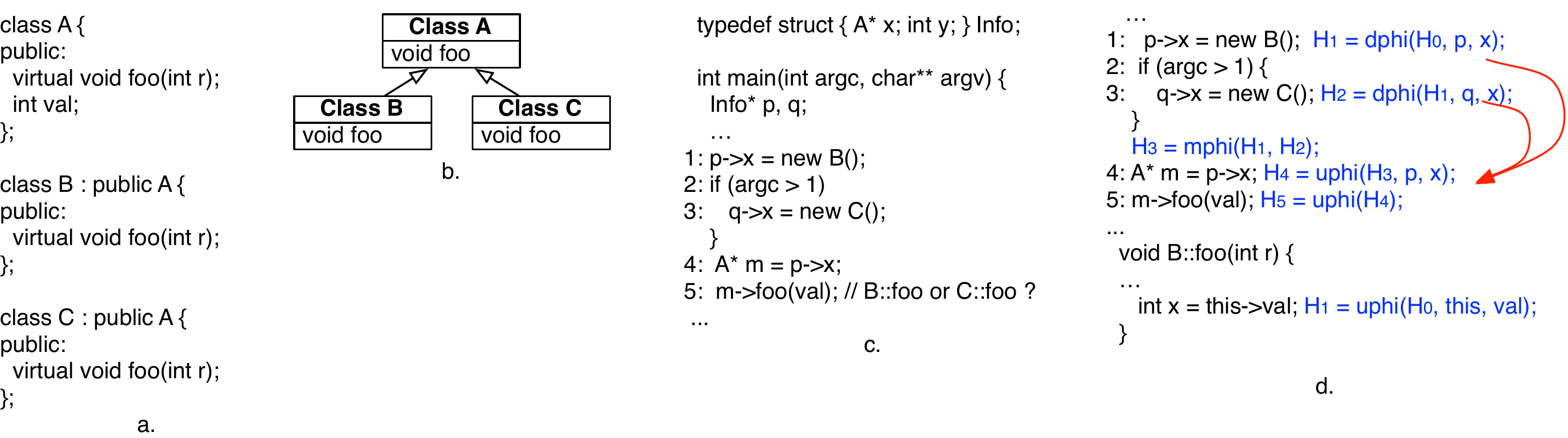}
\caption{The Class Type Analysis Example}
\label{fig:rtas_example1}
\end{figure*}

In C++ code, the analysis needs to identify a minimal set of possible class types in the presence of class inheritance.
This helps the call graph builder to precisely identify the target of virtual function calls. The
first step of class type analysis (CTA) is class hierarchy analysis (CHA), which examines the
class information to build the tree structure that represents the C++ class hierarchy.
Figure~\ref{fig:rtas_example1} (a) gives a simple class hierarchy example, where classes B and C are
subclasses of class A. The class hierarchy tree is presented in Figure~\ref{fig:rtas_example1} (b).

The next step of CTA is to start from class instantiation sites and propagate class type information
via variables' def-use chains. LLVM provides scalar variable-based SSA form to represent def-use 
information for scalar variables. For heap variables, CTA needs assistance from pointer analysis. In 
Figure~\ref{fig:rtas_example1} (c), the example code presents a case where pointer analysis information 
can disambiguate class types. In Line 5, the value of variable {\em m} is loaded from
{\em p$\rightarrow$x}, which can be an instance of class B or C. 
To identify the type of variable {\em m}, we need to know if variables {\em p} and {\em q} are aliased or not. 
If from pointer analysis we know that {\em p} and {\em q} cannot be aliased, then {\em m}'s class type
is B, and the invoked function {\em foo} in Line 6 is {\em B::foo}. Otherwise, both {\em B::foo}
and {\em C::foo} may be invoked at Line 6. 

We now describe an interprocedural, flow- and field-sensitive class type analysis that starts from class 
instantiation sites and propagates class type information via variables' def-use chains. The def-use 
information is built upon both scalar SSA (for scalar variables) and HSSA (for heap variables, see more 
details in the next section). For each scalar variable defined, all of its uses are checked and their class 
types are updated. If the use is a merge $\phi$ function, a meet update operation is performed, i.e. 
merging the class type into the merge $\phi$ function's class type set. For each heap variable defined, 
all of its may-alias uses are checked and their class types are updated (i.e. merging the class type 
into the heap variable's class type set). The operation of the heap variable's merge $\phi$ is the same as for scalar 
variables. 

\subsubsection{Heap Static-Single Analysis Form}
Information flow analysis discovers the flow of values between variables in a given application. The 
variables can be scalar or heap variables. Heap SSA (HSSA) form~\cite{DBLP:conf/sas/FinkKS00} is 
used to represent the definitions and uses of heap variables,
i.e. class/struct field and array accesses in the C/C++ context.
For each heap variable definition and use, a pseudo-variable {\em H$_{i}$} is used to annotate the heap variable 
access, where a {\em d$\phi$} function is used for definitions and a {\em u$\phi$} function for uses. 
The {\em d$\phi$} and {\em u$\phi$} functions take the
heap address (e.g., {\em p}) and offset (e.g., the offset of struct Info's field {\em x}) as input
parameters that represent the heap position. Similar to scalar SSA, a merge {\em $\phi$}
node is used to merge {\em d$\phi$} or {\em u$\phi$} nodes where control flow edges join. 
Figure~\ref{fig:rtas_example1} (d) shows the transformed HSSA form from
Figure~\ref{fig:rtas_example1} (c). 
Two  {\em d$\phi$} functions   (i.e., {\em H$_{1}$} and
{\em H$_{2}$}) are added to heap definitions at Lines 1 and 3, one {\em u$\phi$} (i.e., {\em H$_{4}$})
is added to a heap use at Line 4, and a merge {\em $\phi$} node is used to merge {\em H$_{1}$} and
{\em H$_{2}$}.

Recall from the CTA algorithm, that class type  information can be propagated via HSSA def-use chains.
At Line 1, {\em H$_{1}$} is assigned  class type B. {\em H$_{2}$} is assigned class
type C. {\em H$_{1}$} propagates its class type information through HSSA def-use chains, and
reaches {\em H$_{4}$} as a use, since {\em H$_{1}$} and {\em H$_{4}$} are must-aliases.  
So {\em H$_{4}$} takes on class type B. For variable {\em m} at line 4, its class types depend on the 
type of {\em p} and {\em q}, since the definition of {\em H$_{4}$} comes from {\em H$_{3}$} which 
merges {\em H$_{1}$} and {\em H$_{2}$}. If {\em p} and {\em q} may aliases, then {\em m} takes 
class type B and C. If {\em p} and {\em q} must not alias, then {\em m} takes class type B only. 
Building the HSSA form simplifies the  manipulation of heap variables for analysis. The may/must alias 
checking gets help from pointer analysis or value numbering (i.e., the GVN in LLVM).

For function invocations, HSSA connects those call sites whose target functions have a side effect (i.e. 
a load or store of a heap variable). For example, a {\em u$\phi$} function is assigned to the 
invocation of function {\em foo} at Line 5 in Figure~\ref{fig:rtas_example1} (d), since function 
{\em foo} performs a load operation on heap variable {\em B::val}. 

\subsubsection{Call Graph Construction}
As discussed above, precise call graph construction (CG) for C++ code needs precise class type
information to identify virtual function calls. Based on the CTA analysis output, the CG
builder starts from entry functions.  In this paper, entry functions are the 
{\em main} functions and event handler functions in the Tizen OS code base and mobile applications. 
For each indirect function invocation (i.e., invocation via a function pointer), if it is a virtual function
invocation (i.e., the function pointer is loaded from a class object's virtual table), the target
object's class type information is used to identify target functions. For a non-virtual indirect function  
invocation, CG builder uses pointer analysis information to identify the target functions. 

\subsection{Handling C/C++ Features} 
C/C++ has features that pose difficulties for static analysis, such as the coexistence of C++ inheritance with C function pointer use, 
the coexistence of classes/structs, and array elements accesses in the form of offsets from pointers.  
SIFA's call graph construction, as described above, integrates the handling of invocations through function pointers with the handling of virtual function calls.  
SIFA extends Heap SSA (HSSA) form \cite{DBLP:conf/sas/FinkKS00} to represent memory accesses through class/struct field accesses and arrays  in a uniform way.   
The original work in heap SSA  only supported Java objects, and was extended for C/C++ objects in this work.

\section{Tizen Application Analysis}
\label{sec:analysis_app}
The Tizen application analysis is an interprocedural SIFA analysis that identifies pairs of taint source and sinks for the given application code.
The taint source and sink pairs are defined by user-specified rules, i.e. the taint source function as the key and a set of taint sink functions as values. 
Taint analysis can be used to model different security issues, such as privacy leaks and unauthorized resource access.
Here we focus on privacy leaks.
The analysis engine loads the user-defined taint source and sink map into memory, and analyzes the mobile 
application code (represented as LLVM IR) to identify taint source function invocations. 

Like all data flow analyses, taint analysis defines an associated lattice and meet function. 
The top element of the lattice is \texttt{Untainted}.  The bottom element of the lattice is \texttt{Tainted}.  These are the only two lattice elements. 
A variable definition is assumed to be initialized to \texttt{Untainted}, and becomes  \texttt{Tainted} if it is assigned to by an expression containing a tainted value.
Where a definition is assigned to by a $\phi$-function, it becomes \texttt{Tainted} if any of the arguments of the $\phi$-function are tainted.
Thus the meet operation for the lattice is defined as:  \texttt{meet(Tainted, Untainted) = Tainted}. 
For each taint source function invocation, the tainted value is    
propagated through scalar SSA and HSSA def-use chains.  
When a taint source reaches a corresponding sink function,  then a taint pair
is identified and reported to output. 

\begin{figure}[htbp]
\center
\includegraphics[width=3.0in]{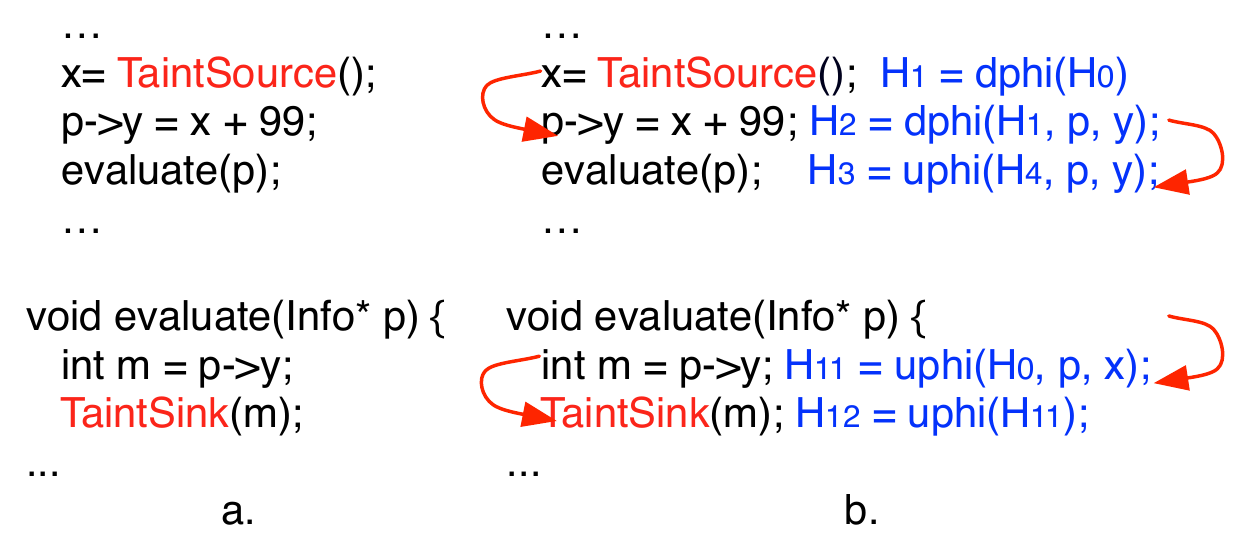}
\caption{Example of Interprocedural Taint Analysis}
\label{fig:rtas_example2}
\end{figure}

Figure~\ref{fig:rtas_example2} (a) shows an example where a taint source and sink are identified across
procedure boundaries. The {\em TaintSource} function is invoked and produces the result value {\em x}
that should be marked as tainted. By propagating through dataflow analysis, all variables in the 
computation reached by the tainted value are marked as tainted. In the function {\em evaluate}, the 
sink function {\em TaintSink} is invoked and has tainted variable {\em m} as input. Thus the taint 
source reaching its corresponding taint sink is identified. To illustrate the dataflow traversal, 
Figure~\ref{fig:rtas_example2} (b) gives the HSSA version of the code, and the arrow lines show the 
taint lattice value propagation through the scalar and heap variable def-use chains. 

We perform taint analysis in time that is linear in the size of the HSSA graph. 
The implementation is currently context insensitive.

\subsection{Implicit Flows} 

Our static taint analysis, unlike existing tools (~\cite{Sridharan:2007,Tripp:2009,Arzt:14}), identifies implicit flows~\cite{King:2008}  due to control dependences between (source, sink) pairs.
This is needed to ensure that a malicious program cannot sidestep the
taint flow policy rules through tricky conditionals and control flows.
Our method integrates control-based and dataflow propagation for taint analysis.

For each function, the analysis tracks implicit flows by identifying control predicates and the statements that are control dependent on them. 
A prepass inserts pseudo-uses of the control predicate for each such definition, effectively turning the control dependence relation into a dataflow relation through which the analysis engine propagates taints.  
If the control predicate is tainted, the taint analysis classifies all variable definitions control dependent on the predicate as tainted. 
To implement this, control predicates are inserted as pseudo uses in each conditional statement prior to the taint analysis. 

Consider the code example in Figure~\ref{fig:rtas_example4}:

\begin{figure}[htbp]
\center
\includegraphics[width=2.0in]{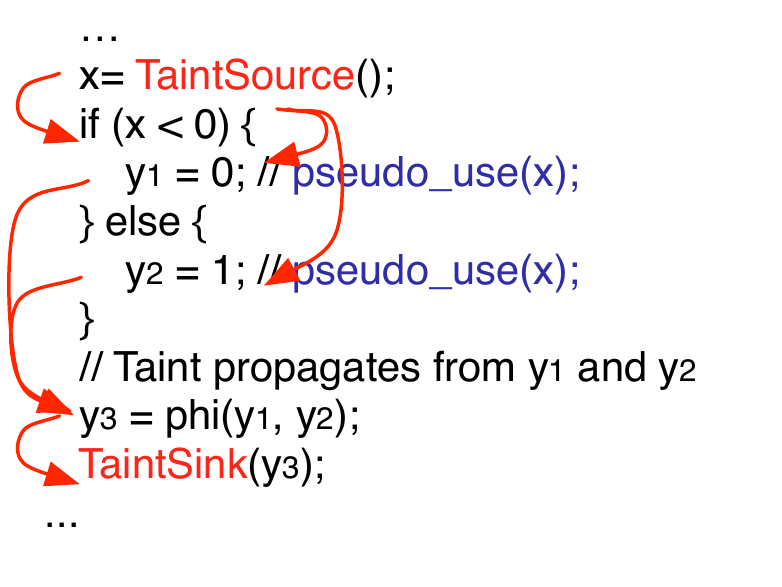}
\caption{Example for Pseudo-Use}
\label{fig:rtas_example4}
\end{figure}

In the code example, the tainting of $x$ is propagated to $y_1$ and $y_2$ through the insertion of pseudo uses. 
Thus $y_3$ is tainted, and so a privacy leak occurs at 
\texttt{TaintSink($y_3)$}.

\subsection{Input Rules}

Taint rule specification is usually done by identifying sources and sinks at the API level, but this may lead to unnecessary loss of precision, especially for languages such as C and C++ that need to account for reference parameters, pointer parameters and inheritance.   
We allow for a more refined specification in which the source and sink are identified as specific API parameters (including return values) of APIs. 
For example,  it is possible that a security analyst may consider {\tt image}, but not {\tt metadata}, to be
a taint source in an API call like {\tt GetImage(\&image,   \&metadata)}.   
Likewise, {\tt filename}, but not {\tt mode}, may be
considered to be a taint sink in an API call like {\tt OpenFile(filename, mode)}.

\subsection{Callback functions}

Callback functions pose an interesting challenge because they can
enable ``hidden'' information flow via event-driven execution.
Consider for instance, the snippet of code (shown in Figure~\ref{fig:rtas_example5}) in which an
application uses a callback function to preview a snapshot captured by the camera device:

\begin{figure}[htbp]
\center
\includegraphics[width=2.8in]{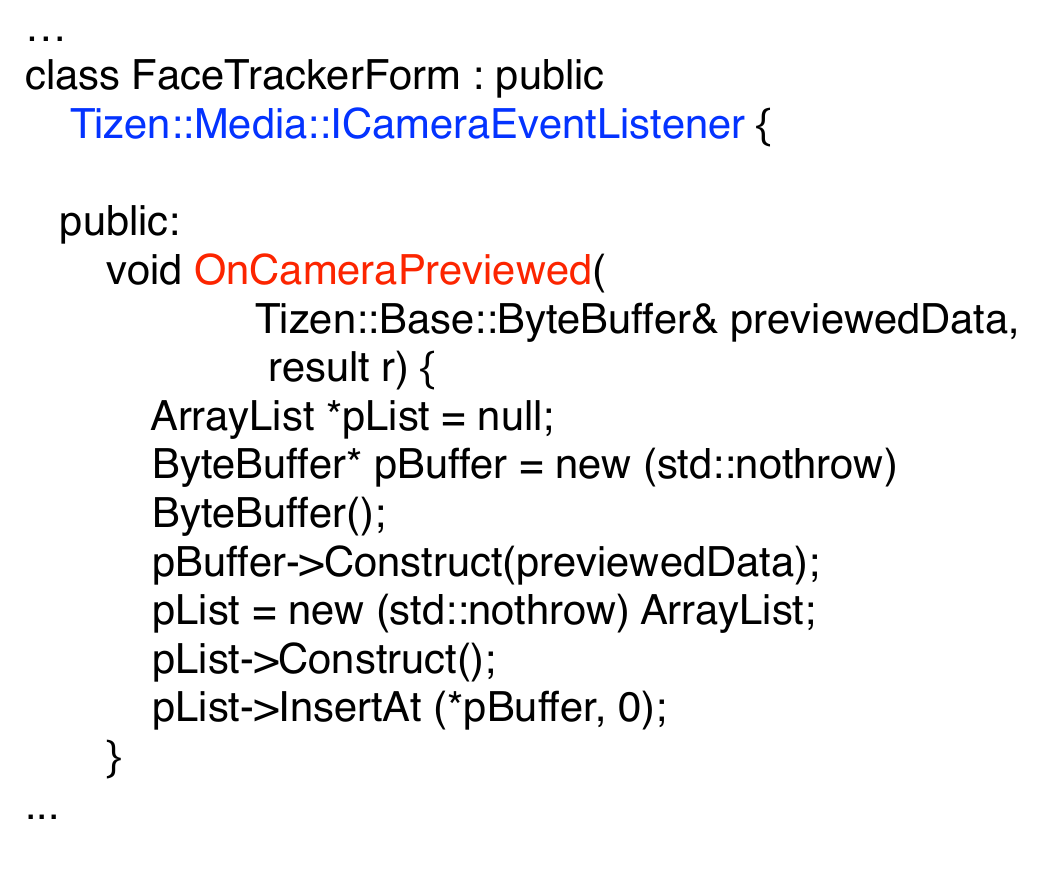}
\caption{Example for Event Handler}
\label{fig:rtas_example5}
\end{figure}

In the example above, the \texttt{FaceTrackerForm} class implements an interface function 
called \texttt{OnCameraPreviewed()} exposed by the 
\texttt{Tizen::Media::ICameraEventListener} class. The \texttt{ICameraEventListener} 
class is meant to provide callback functions to retrieve data in an event-driven fashion. 
In the example above, the \texttt{OnCameraPreviewed()} function retrieves the captured 
snapshot packaged as a (Tizen) \texttt{ByteBuffer}.

To handle callback functions in our taint analysis, the input rules make use of a 
\texttt{type} attribute, where for callbacks the type is set to \texttt{event}.
These extra attributes compensate for our desire not to include the
entirety of the Tizen system libraries as part of our information flow
analysis of potentially hostile apps. Instead, we only need to
annotate the various library entry points and their callback behaviors.

\subsection{Ranking of Vulnerabilities}

The output of the taint analysis is a prioritized list of vulnerabilities.  
Each vulnerability rule assigns a severity level. 
For instance, leaking one's location might be considered to be a lower severity than leaking one's SMS messages. 
The vulnerabilities detected by taint analysis are ranked primarily
according to their severity and secondarily according to the distance
between source and sink in the application, under an assumption that a
longer distance from source to sink represents less of a security
threat.   
Of course, all of this is still reported to the human analyst.

Where vulnerabilities have the same severity level, their relative ranking is based on their distance metrics. 
A shorter distance results in a higher rank. 
The attributes  \texttt{$call\_distance$} and \texttt{$control\_distance$} 
define a metric for the distance between the source and the sink in the application. 
The $call\_distance$ value is one if the path from the source to the sink includes a function call and is zero otherwise.
Where the source dominates the sink in the control dependence graph, the control distance is the length of the path between them. Where the source does not dominate the sink, the control distance is the sum of the distances between each node and their least common ancestor. 
The $control\_distance$ value is only relevant when the value of $call\_distance$ is zero.

As in \cite{Kremenek:03}, the ranking of the vulnerabilities can be used to sort them so that the most likely errors appear closer to 
the top of the vulnerability list generated by the taint analysis. 
A tunable cutoff threshold (e.g., top 100) 
of  the vulnerabilities can be included in the output report. A smaller threshold will decrease the false 
positive rate but increase the false negative rate.

\subsection{Tizen Application Analysis Evaluation}

\subsubsection{Tizen Application Analysis Results}

We wrote a rule set for Tizen application analysis, based on Tizen security policies. Using our tool, we were able to find unexpected behavior for an application. 
	
We used 30 Tizen sample native applications, which were the only available applications during the time of this research. 
We created rules to detect privacy leaks and unauthorized resource accesses involving the file system.  
These are among the security vulnerabilities that TizenStore.com would check for to ensure the safety of Tizen apps prior to being downloaded to Tizen users.
For both cases, rules consist of one taint source API and one or more taint sink APIs.  
We also checked colluding apps, which needs support for identifying taint pairs cross IPC cals. 
For this case, we created two rule sets for 
colluding applications (one for ``producer'' applications with information flow from the SMS to IPC calls), 
and one for ``consumer'' applications with information flow from IPC calls to File).  
The two rule sets are marked that the analysis engine can recognize them and apply them as producer/consumer pattern,

Our tool identified one privacy leak in the {\tt FriendFinder} application without any false 
positives or false negatives. In the {\tt FriendFinder} application's {\tt ConnectionManager} class, there is a function 
{\tt GetImagePathPtr} that retrieves the path information as a string. In the same function, there is a 
{\tt BluetoothOppClient::PushFile} function that takes the output string of {\tt GetImagePathPtr}. 
This induces a privacy leak because the {\tt GetImagePathPtr} API is obtaining a profile picture 
(i.e., file name) of the user and sending it to another device via the 
{\tt BluetoothOppClient::PushFile} API.  

With a finding like this, an analyst looking at this report might
conclude that FriendFinder is operating as expected, sending profile
pictures through the Bluetooth connection would seem to be an expected
behavior for the app. If there were a flow to the network, however,
then the analyst would have reason for concern and might take action
to ban the app.

\subsubsection{Tizen Application Analysis Performance}

We ran SIFA on a quad-core Intel Xeon 2.66GHz workstation with 8GB of memory and running RedHat Linux (RHEL 5). 
The largest application is {\tt MediaApp}, which contains 129,375 bitcode instructions. 
{\tt MediaApp} took the longest time to analyze: it took 22.82 seconds for total execution. 
The analysis, which includes the related LLVM analysis pre-passes, pointer analysis, and interprocedural taint analysis,
took 20.351 seconds. Our experiment shows that the tool can consume more than 10,000 LLVM bitcode 
instructions per second (i.e., about 3,000 lines of C++ code per second) on average. We also measured peak memory usage using 
Valgrind and the largest memory consumption came from {\tt MediaApp}, which required 3.098GB. 

\begin{figure}[rtas1throughput]
\center
\includegraphics[width=3.0in]{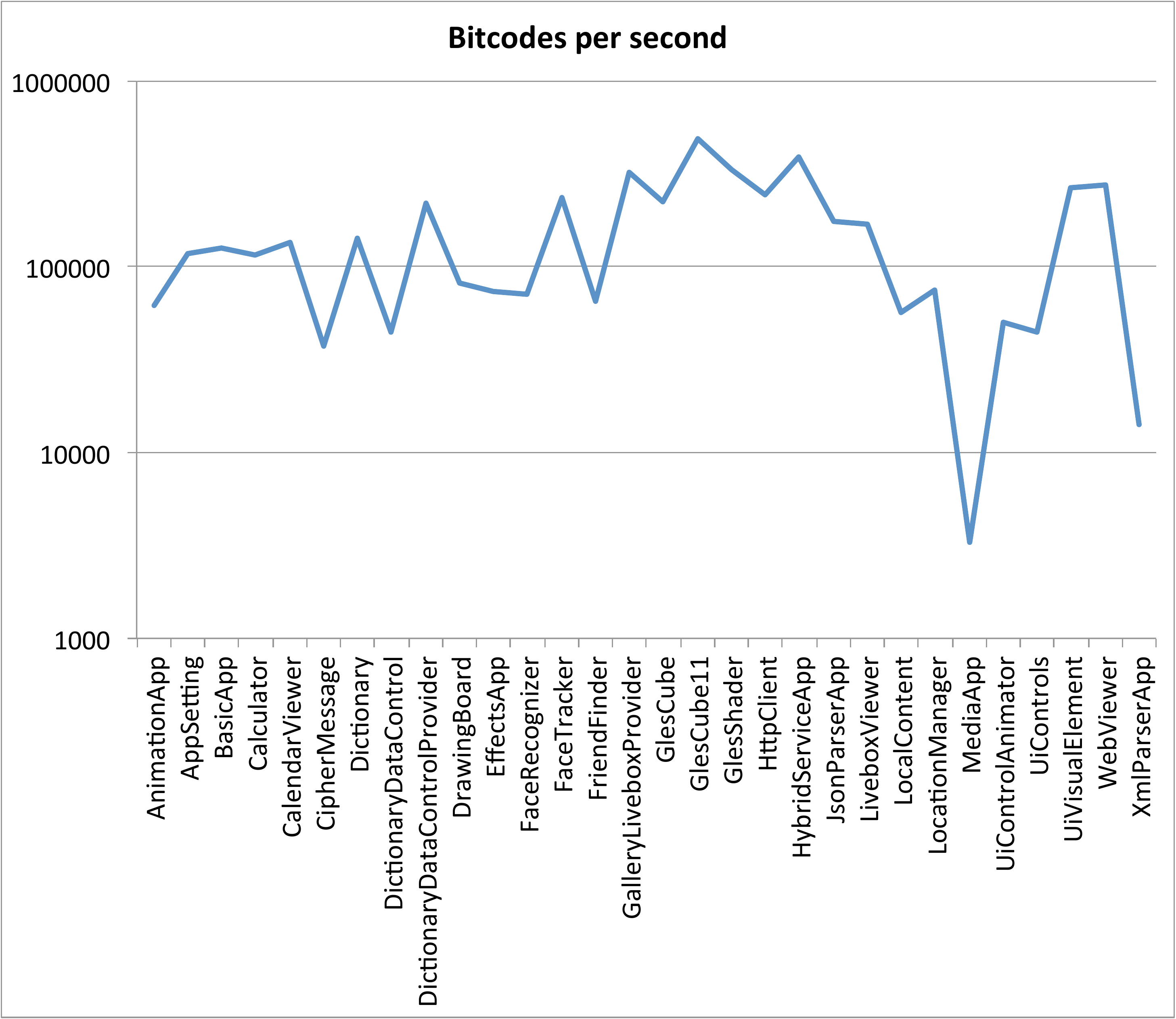}
\caption{Tizen Application Analysis Throughput}
\label{fig:rtas1throughput}
\end{figure}

For contrast, we note that Google's Play Store for Android introduces
several hours of latency between when an app is submitted and when it
becomes live for production. The CPU and time costs for performing our
information flow analysis are negligible compared to the time a human
analyst might spend understanding them and considering whether the
results are appropriate for the app's claimed functionality. And, of
course, as the volume of submitted apps grow, standard cluster
resources can be used to conduct concurrent analyses, independently, 
with human analysts engaging after the analyses are complete.
 
\section{Tizen API Analysis} 
\label{sec:analysis_api}
Tizen API analysis (TAA) identifies paths from native API calls to low-level system (Linux) kernel calls to
test for potential violations of user privileges. It performs a dataflow analysis on top of the call graph
to identify information flow. 
Here the propagated information is the set of user privileges exercised along call paths.

The user-specified privilege rules are inputs to the analysis, defined as: 
\begin{enumerate}
\item A set of (source, sink) pairs, where each source is a native API call and each sink is a glibc call, which is a wrapper for a kernel call;
\item A set of user privilege properties (UPVS) that call paths from the source to the sink, 
for each (source, sink) pair.
\end {enumerate}

TAA traverses the call graph in a top-down manner from each entry function (the call graph here is a forest), and starts a new call path
trace when a {\em source} call is identified. 
An entry function here is an event handler function in the Tizen OS code base.  
The call path trace is performed on the call graph by means 
of HSSA. For each path in the library code base, the TAA collects the set of user privileges (PVS)
exercised along the call path and stored the call path into a candidate list when a {\em sink} call is identified. 
The privilege is checked from a {\em CheckUserPrivilege} function call
but the user can also specify other special function calls for identifying user
privileges. 
A call path is a potential violation of user privilege properties, iff its PVS contains an element that is not
in UPVS (i.e. PVS is not a subset of UPVS).

\begin{figure}[htbp]
\center
\includegraphics[width=3.2in]{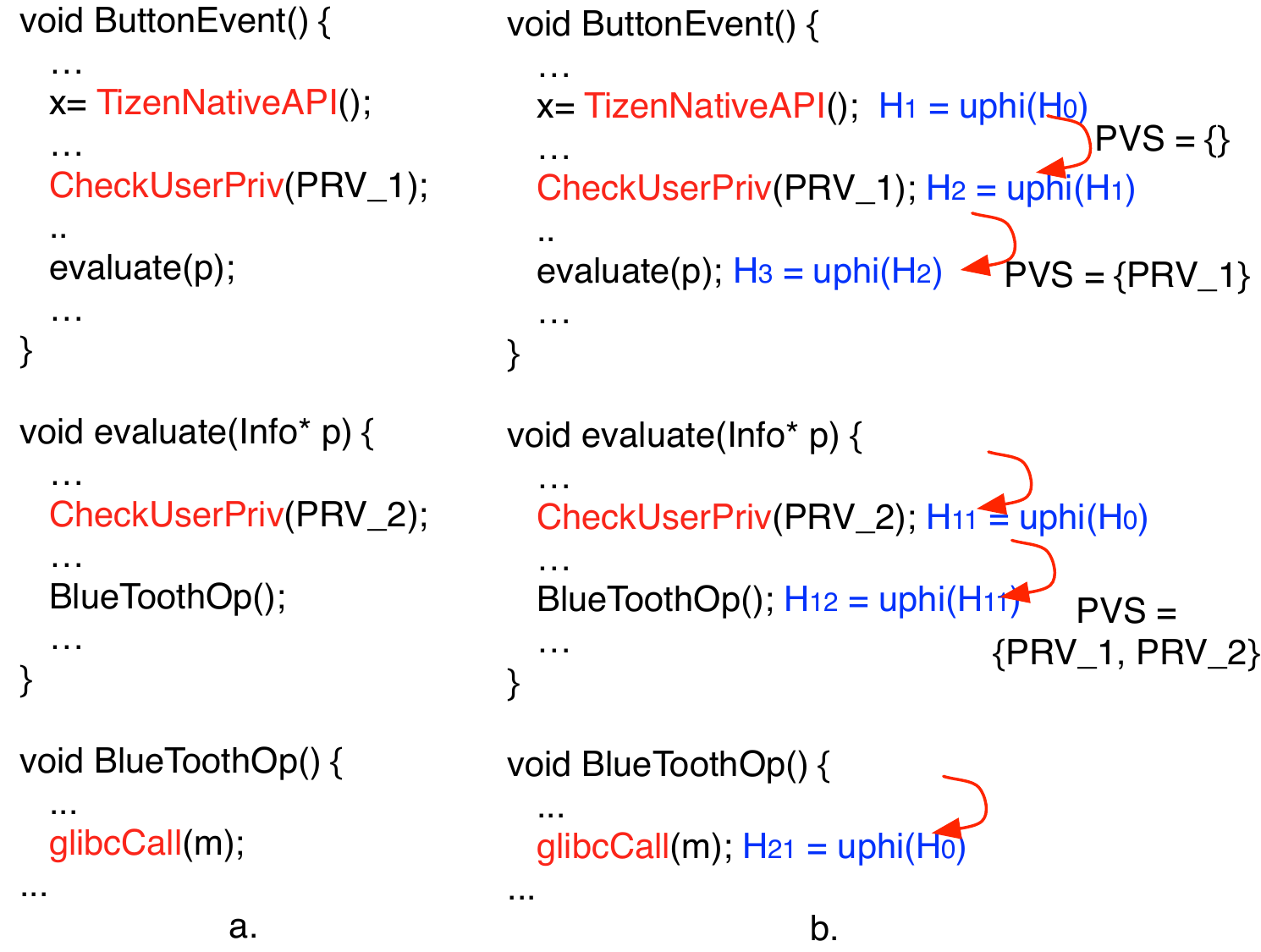}
\caption{Example of Tizen API Analysis}
\label{fig:rtas_example3}
\end{figure}

Figure~\ref{fig:rtas_example3} (a) shows an example where the code base contains a user-specified {\em source}:
{\em TizenNativeAPI}, {\em sink}: {\em glibcCall}, and the check privilege function: {\em CheckUserPriv}. 
There is a call path from {\em ButtonEvent} $\rightarrow$ {\em evaluate} $\rightarrow$
{\em BlueToothOp}. There are two user privileges exercised in this call path: {\em PRV\_1} and {\em PRV\_2}.
Figure~\ref{fig:rtas_example3} (b) gives the HSSA version of the code (only function based {\em u$phi$}
nodes need to be considered in this analysis), and the arrow lines show the progress of updating PVS in the
HSSA def-use traversal for call path.

The output of TAA is a list of such call paths that potentially violate user privilege
properties. The output includes the source Tizen API function, the sink Linux kernel function,
and the full call path from source to sink. This analysis can lead to false positives and false
negatives. Since this SIFA runs on API library, it can be extended to model additional security
issues, such as unauthorized resource access.

\subsection{Tizen API Analysis Evaluation}

We wrote a rule set for Tizen API analysis based on the Tizen security policies discussed above. Using our tool, we were able to find several unexpected behaviors of the Tizen APIs.

\subsubsection{Tizen API Analysis Results}

For the Tizen API analysis, we started with a simple rule to test whether Tizen enforced privilege checks for the privileged APIs. 
Our tool located a privileged API which didn't follow the API documentation~\cite{TizenAPI}. This bug allows applications to receive push notifications without owning one of the two required privileges. 
While this API requires {\tt \_PRV\_PUSH} and {\tt \_PRV\_HTTP} according to the API documentation, it only checks for the {\tt \_PRV\_PUSH} privilege. Our tool detected this inconsistency.

Furthermore, we found two API calls which have the functionality of registering an application to the application launcher so it can run when a specified condition is met (comparable to an Android app's ability to register to receive a broadcast intent).
 One API call can register {\em any} application while the other can only register its caller. Our tool detected that the broader API call is vulnerable in that it doesn't have a required privilege check while the other API has it. Thinking we found a significant vulnerability, we dug deeper and followed the subsequent execution path manually. We ultimately discovered that the app launcher, itself, which receives these calls makes its own security checks. 
 While this finding could be interpreted as a false positive, the discrepancy between security checks taking place on different levels for the same mechanism is something that deserves manual scrutiny. Our tool allowed us to focus our attention on an API call that indeed appeared to have an exploitable hole.

Our analysis also highlighted several InputMethod APIs. None of the InputMethod's privileged APIs had privilege checks, including the SendText API. Again, we manually followed the calls and discovered that, unlike other classes' privilege checks, InputMethod enforced privilege checks in GetInstance when the application retrieves an instance of InputMethod. In this respect, InputMethod follows something of a capability-style of access control (i.e., if you hold a valid instance, then you must be allowed to use it). So, while we again didn't find a vulnerability, we did find a coding style at odds with the way the rest of the APIs do their security checks, deserving of additional scrutiny.

Lastly, we wrote another rule that detects flows from the privileged APIs to non-privileged APIs. The intuition behind this rule is that if a privileged API only uses non-privileged APIs, the privilege check is unnecessary. We found a privileged API which deletes all cookies in an application that could be replicated only using non-privileged APIs. While this doesn't indicate a security hole, it does validate that our tools is capable of discovering both missing security checks as well as unnecessary ones.

Overall, while we're modestly disappointed that we didn't find any security flaws, we note that a massive codebase like Tizen, with a large stable of developers contributing new code on a regular basis, creates logistical challenges for the security analysts trying to keep up with it. A tool like ours, running as part of a nightly build system, allows an analyst to detect new flows and control paths that might have innocently introduced security vulnerabilities. 

\subsubsection{Tizen API Analysis Performance}

Our analysis ran on a quad-core Intel Core i7-3770 3.50GHz workstation with 8GB of memory, running 
Fedora Linux and LLVM 3.3. The test bed is a part of the Tizen platform consisting of 4,346 C/C++ files compiled into 
LLVM bitcode files with a total size of 560MB. The analysis time for generating the call paths for all APIs took 122.5 secs with memory usage under 8GB. This is fast enough that it could reasonably run not only as part of a nightly build process but as part of a regular developer's source code commit process, flagging new flows before the change hits the code repository.

\section{Pragmatic Issues}\label{sec:pragmatic}
Our static analysis tool leverages the LLVM analysis infrastructure and so depends on the use of the LLVM compiler. For the Tizen native application analysis, LLVM/Clang is the default compiler. However, the Tizen platform code is compiled using GCC. To compile the Tizen platform code with LLVM,  we had to address issues that other large-scale static analysis tools --- such as Coverity --- also had to address when processing real-world software: the issues raised by standards, language dialects and compiler variations~\cite{Coverity}. In short, to use the LLVM infrastructure, we had to make two changes to the Tizen source distribution. 

First, we needed to change the compiler from GCC to Clang, which generates the LLVM bitcode that is input to the LLVM analysis infrastructure. Since GCC and Clang are not completely compatible~\cite{ClangVsGcc}, this step involved manual inspection of each module. 
We edited each build file and made source code changes as needed to remove errors. Changes, in some cases, included editing of assembly code. 

Second, Tizen uses a variety of different build systems (CMake, libtool, and traditional makefiles). Consequently, each module is a new adventure in software porting, both in terms of the initial compilation step and as well in terms of linking. 

Consequently, we had to decide when we had enough coverage to validate our tool and approach. 
The Tizen source is divided into different source packages and we successfully compiled 159 out of 390 Tizen framework packages to LLVM bitcode, generating more than 4,000 LLVM bitcode files with a total size of 560MB. We compiled all the packages from the top two layers: OSP and the CAPI layer, which handles the native application.  
We picked underlying components' packages that were directly relevant to the privileged APIs such as telephone, messaging, system, and etc. We did not compile packages that were not relevant to the privileged APIs such as graphics, UI, and multimedia.

A full analysis, of course, would need to push the entirety of the Tizen codebase through LLVM, and this effort would need to be replicated each and every time the analysis was to be conducted. If our vision of our tool being closely integrated in the Tizen build environment were to ever take off, Tizen would realistically need to switch to LLVM as its production compiler. With LLVM in production use by a number of very prominent projects, include Apple's iOS / OS X, this isn't an unreasonable recommendation.

\section{Related Work} \label{sec:related}
\subsection{Static analysis of production code}
Static analysis has been proven to be successful in finding bugs in real-world programs. Coverity~\cite{Coverity} and Fortify~\cite{Fortify} are well-known commercial static analysis tools. An article by Bessey, et al.~\cite{Coverity_issue} discusses a number of pragmatic issues and experiences with respect to static analysis tools for finding bugs for large  commercial code bases (up to  20-30 MLOC). They observe that "the false positive rate is simplistic since false positives are not all equal and initial reports matter inordinately". 
Both Fortify~\cite{Fortify} and Coverity emphasize results prioritization once vulnerabilities are identified. 
Our ranking and cutoff analysis (Section~\ref{sec:analysis_app}) also addresses this issue. 
We discuss related ranking work in Section ~\ref{subsec:related_ranking}. 

IBM AppScan Source~\cite{AppScan} is a tool meant to identify bugs during the development phase for web applications.  Other editions of IBM AppScan identify general bugs while focusing on security problems in particular and supporting  customizable rules. 

FindBugs~\cite{Ayewah:10}, a static analysis tool used on Google code bases, focuses more on identifying common Java programming bugs rather than security vulnerabilities in particular. The importance of the tool's UI with respect to the speed of understanding and fixing bugs has been demonstated~\cite{FindBugsFaster} (analysts processed bugs in FindBugs faster than with Fortify). The tool was used to show that bugs found in older code bases are less likely to be fixed once discovered~\cite{FindBugsOlder}.

ESC/JAVA~\cite{ESC_JAVA} is a static analysis tool, powered by verification-condition generation and automatic theorem-proving techniques, for Java that checks for common programming errors.  While it does find errors, users have to annotate the software and the annotation burden is quite high. It also suffers from excessive spurious warnings on programs that are annotated.  

Metal~\cite{Metal} is a language for programmer-written compiler extensions that express a broad range of correctness rules that code must obey.  
The system  xgcc executes these extensions using a context-sensitive interprocedural analysis. 
Metal is designed for system programmers with an emphasis on ease of use, and makes use of state machines as a fundamental abstraction. 
This approach has been used to find thousands of bugs in real systems code.

\subsection{Security analysis of mobile applications}
Privilege escalation attacks on mobile applications are known to the community. In particular, 
the vulnerability of Android applications~\cite{privilegeEscalation} is well known. 
Android, like Tizen,  is a permissions-based mobile operating system, so analysis of possible permission leak vulnerabilities is also needed for it. 

ScanDroid~\cite{scandroid} was the first static analysis tool for Android to detect 
information flow violations. The tool detects inter-application security risks and needs to have 
access to both the vulnerable application and the exploitable application. To the best 
of our knowledge, SCanDroid is not easily extensible with new taint propagation rules, unlike SIFA
which is designed from the ground up for supporting custom rules.

FlowDroid~\cite{Arzt:14} is a static taint-analysis tool for Android applications, based on the Heros FDS/IDE solver and the Soot Java analysis framework. 
It models the Android application life cycle, including multiple entry points, asynchronously executing components, and callbacks.   
It performs context-, flow-, field-, and object-sensitive analyses to discover vulnerabilities in applications. 
FlowDroid has excellent performance because it performs on demand alias analysis, but as described in FlowDroid~\cite{Arzt:14} it does not handle implicit flows through control dependences. 

Grace et al. \cite{stock} focus on static analysis of stock Android firmware and 
identify confused deputy attacks that enable the use of permission-protected capabilities. 
Our application analysis is complementary in that it identifies not only actions that are performed, but 
information that flows to attackers. Our focus is not on stock applications, but on third-party 
applications.

CHEX~\cite{Lu:2012}, relies on taint analysis to discover permission leaks in Android 
applications. CHEX detects several types of vulnerabilities affecting Android applications, 
including permission-protected information leaks. The CHEX analysis is similar to our application analysis, but relies on a model of the OS libraries rather than analyzing them directly. This avoids handling the multi-language analysis difficulties that Tizen and Android have.

TaintDroid~\cite{taintdroid} uses dynamic taint tracking to identify protected information flows 
that reach Android network communication APIs (sinks).
Advantages of performing this analysis dynamically are increased precision, as well as enforcement of the safe use of vulnerable applications by denying users the capability to externalize their sensitive information during application use. The advantage of static analysis tools such as SIFA is their capability of detecting vulnerable applications before they even reach the user.

ComDroid~\cite{comdroid} is a tool that analyses inter-application communication in Android. 
ComDroid does not track permission-leak vulnerabilities and none of the discovered vulnerabilities described 
pertain to permission-protected information. Contributions such as automatic rule generation 
separate our work from theirs. 

Kirin~\cite{NASTR, Enck:2009} is a tool based on a formal representation of the Android security 
model that checks if applications meet security policies. It can check for confused deputy 
vulnerabilities (``unchecked interface''),  Intent spoofing (``intent origin'') and other attacks 
by using a powerful Prolog-based security policy enforcement mechanism, which takes into 
consideration the set of applications already installed on a device. The authors point out 
several difficulties with creating information flow policies in Android
and discuss the future possibility of including source code analysis to make information 
flow policies for Android of practical use. 

Felt et al.~\cite{demystified} map Android API calls to permissions
based on automated testing rather than static analysis, which means incomplete 
coverage and the possibility of false negatives in the permissions map. They do not use the map to 
check for information flow-based vulnerabilities in applications.
PScout~\cite{Au:2012} builds a permission map for Android through static analysis based on Soot. 

A different aspect of the flow vulnerabilities is described by Claudio Marforio et al., whose work focuses on colluding applications~\cite{2011_application_collusion}. 
They identify  several possible covert channels through
which malevolent applications can communicate sensitive information, for example by enumerating 
processes using native code or files. Most of these however are not Android specific. 
They did not build a tool to detect flow vulnerabilities. 
They identify security risks for colluding applications in modern 
permission-based operating systems.

The PermissionFlow tool ~\cite{Sbirlea:13} performs a static dataflow analysis to identify sources of information protected by permissions in Android and a taint analysis to check if this information reaches other applications or leaks outside the device. The source APIs are, in contrast to the work of Felt et al., obtained through static analysis. PermissionFlow does not offer any support for implicit flows.  Bartel et al.~\cite{Bartel} propose a similar taint analysis and both were able to find vulnerabilities in commercial applications, highlighting the importance of performing a corresponding analysis on Tizen applications. --- which is what our tool does.

\subsection{Results Ranking}\label{subsec:related_ranking}
To the best of our knowledge we are the first to use \textit{error ranking and cutoff} as means to reduce the false positive rate in security analysis, but there is a long history of using these techniques in static analysis tools. 
We use error ranking to both suppress false positives and to prominently display errors that are considered to be of importance to the user. 
The goal of \textit{report prioritization} is to display errors according to their importance to the user.  
An article by Bessey, et al.~\cite{Coverity} observes that the most prominently displayed reports are critical and have a strong impact on the user's perception of the quality of the tool. 

An early tool to use error ranking for results of a static analysis is Prefix~\cite{Bush:00}, which focuses on analysis of memory allocation and usage errors in C.
It was an essential tool for improving the reliability of Windows OS\cite{Das:06}. Because of the high volume of warnings generated, Prefix uses a set of ad hoc filters to improve the relevance of the warnings displayed.

Several tools, such as Z-Ranking~\cite{Kremenek:03}, Feedback-ranking~\cite{Kremenek:04} and Airac~\cite{Jung:05}  propose the idea of using statistical modeling to obtain better ranking of positives.  Kremenek observed~\cite{Kremenek:04} that bugs often cluster by code locality and attributes this characteristic to the observation that programmers who do violate a particular programming rule tend to violate it multiple  times.  
Code locality plays a role in our ranking function too, but the correlation is instead between the confidence in a result being a  true positive and the code span of the taint.

FindBugs~\cite{FindBugsOlder} performs report prioritization by combining several factors such as confidence in a result and the seriousness of the bug. In our system, report prioritization is accomplished by sorting security violations 
according to severity. As with our work, the severity is provided based on user-specified values in the input taint rules. EspX~\cite{Hackett:06} classifies bugs in different buckets based on both its confidence in the error being a true positive and on the severity of the bug.
Fixing all bugs in designated buckets was a requirement to integrate code in the Windows OS~\cite{Das:06}.

\subsection{Implicit Flow}
In this section, we compare SIFA with other work in the area of analysis of implicit flows. 
The implicit flows considered here only include control flows and not covert channels which in general cannot be secured with software-only approaches.
The importance and difficulty of handling implicit flow is presented in numerous studies~\cite{King:2008,Cavallaro:2008}.  
The detection of implicit flows by either static or dynamic analysis has proven to be challenging. 
We have developed a static taint analysis that unifies implicit and explicit information flows in a single analysis mechanism. 

Liu and Milanova~\cite{DBLP:conf/csmr/LiuM10} develop a context-sensitive interprocedural static information flow inference analysis which performs security type inference. 
A security type system requires the annotation of variables and statements with security types, which are labels that denote security levels~\cite{DBLP:conf/popl/Myers99}.  
They handle both explicit and implicit flows. 
Their method captures control dependences through adding implicit flow edges and paths, some of which are annotated by the analysis.
They perform a static taint analysis based on this representation. 

Genaim and Spoto~\cite{DBLP:conf/vmcai/GenaimS05} present an information flow analysis for both explicit and implicit flows for full (mono-threaded) Java bytecode. 
They build a control flow graph that represents the complex control features of Java bytecode.
For efficiency, they represent information flows through Boolean functions.  
They treat fields as static (i.e., global) class variables, and so do not distinguish flow between the same field of multiple objects of a given class, a significant source of imprecision. In contrast, we model objects using Heap SSA form, and so distinguish between the instances of the same field in different objects. 

While taint analysis is effective for detecting a wide range of attacks on benign software, Cavallero, et al.~\cite{Cavallaro:2008} show that it is not as effective for detecting attacks due to malicious software. In particular, they present simple and powerful evasion techniques, used in untrusted x86 binaries, that elude static and dynamic taint-tracking techniques. 
They report that enhancing taint analysis to reason about control dependences, as our method does, improves evasion resistance but results in a high rate of false positives. 
This could limit the usefulness of such techniques, given the wide use of binary-based software distribution and employment models. 
This difficulty motivates the use of trusted LLVM bitcodes as a distribution format.   

King, et al.~\cite{King:2008} experimentally investigate the value of tracking implicit flows through the security-typed language JLift, an extension of Jif. 
They find that implicit flow checking can be valuable, in terms of identifying true leaks of secret information, but produces a high (83\%) rate of false positives (over-tainting), in particular due to unchecked exceptions. 

It has been shown that purely dynamic techniques cannot detect certain implicit flows~\cite{DBLP:conf/sas/Volpano99}, so the application of dynamic taint analysis to implicit flows results in false negatives.  
However to mitigate the issue of static over-tainting,  dynamic taint analysis has been applied to implicit flows through techniques that selectively propagate taints along a targeted subset of control dependences~\cite{Kang:2011,DBLP:conf/issta/BaoZLZX10,DBLP:conf/issta/ClauseLO07}.  
Our use of ranking and cutoff can also be used to mitigate over-tainting.

\section{Conclusions and Future Work}\label{sec:conc}
Analyzing the security of a large software platform like Tizen
presents a valuable opportunity to apply state-of-the-art
tools in static analysis. Static analysis can be usefully applied to
identify undesirable behaviors in apps distributed through app stores,
and it can help the system's developers find needle-in-a-haystack bugs
throughout their system.  While we had a limited library of apps to
consider, we were able to achieve very good analysis performance and were able
to identify non-trivial information flows that could be dangerous in
untrusted apps.  Similarly, by processing a substantial
fraction of the Tizen codebase, we were able to identify a handful of
locations where important security checks were missing; subsequent
manual analysis determined that subsequent software layers made checks
that prevented these initial mistakes from becoming exploitable.

Our work additionally demonstrates the value of a general-purpose infrastructure
like LLVM. While this project focused on C and C++ code, our analyses
could potentially run on any programming language for which there's an
LLVM front-end. For example, a JavaScript front-end for LLVM would
allow our tools to analyze ``web apps'' in addition to ``native apps''
with identical information flow rules.

Furthermore, the extensions we made to LLVM, such as our class
hierarchy analysis, are general-purpose and could well be folded back
into the LLVM distribution. (We intend to make an open source release
of our extensions.)  We hypothesize that the increased precision of
our analyses will enable dynamic dispatches to be replaced with static
function calls, as well as allowing for better function inlining and
other performance benefits. Evaluating this performance impact
represents future work.

Now that Samsung has shipped its first Tizen products and real apps
are starting to appear in its online app store, we expect that
independent security analysts will be able to download these apps, in
bulk, and analyze them as many security analysts have already done for
Android and iOS. The Tizen platform is still in its early days as a
consumer product, creating opportunities for the platform's security
features to get ahead of attackers.

{\footnotesize \bibliographystyle{acm}
\bibliography{bibliography,plas_bib,android,ranking}}

\end{document}